\documentstyle[12pt,epsf,amssymb]{article}
\begin{document}
\setlength\textheight{8.75in}
\newcommand{\be}{\begin{equation}}
\newcommand{\ee}{\end{equation}}

\title{Quasi exactly solvable (QES) equations with multiple
algebraisations \footnote{Talk presented at the 12th Colloquium ``Quantum
Groups and Integrable Systems'', Prague, June 12-14, 2003}}
\author{{\large Yves Brihaye} \\
{\small Facult\'e des Sciences, Universit\'e de Mons-Hainaut, }\\
{\small B-7000 Mons, Belgium }\\
{ } \\
   {\large Betti Hartmann}\\
{\small Department of Mathematical Sciences ,
University of Durham}\\
{\small Durham DH1 \ 3LE , United Kingdom}}

\date{\today}
\maketitle
\thispagestyle{empty}
\begin{abstract}
We review three examples of quasi exactly solvable (QES) Hamitonians
which possess multiple algebraisations. This includes
the most prominent example, the Lam\'e equation,
as well as recently studied many-body Hamiltonians
with Weierstrass interaction potential and
finally, a $2\times 2$ coupled channel Hamiltonian.
\end{abstract}
\medskip
\medskip
\section{Introduction}     
Quasi exactly solvable (QES) operators  can be formed
by considering elements of the enveloping algebra
of some  Lie Algebras realised in terms of differential
operators preserving a finite dimensional vector
space of smooth functions. Some of them can be transformed
into Schr\"odinger operators after performing a suitable
change of variable and a suitable change of function (also
called a ``gauge transformation").
These Schr\"odinger operators admit a finite number of algebraic
eigenvectors and are thus called QES.
For certain QES Schr\"odinger operators  there exist several
choices of the ``gauge functions" which then lead to different
algebraisation of the corresponding eigenvalue equation.
If this is possible, we speak of  ``multiple algebraisations".
The generic example of such an equation is given by the Lam\'e
equation. In this contribution, we present further examples
of QES equations which have multiple algebraisations.
These are equations
describing  N bodies interacting through a Weierstrass potential
and a system of coupled QES equations related to the
sphaleron stability in the Abelian-Higgs model.

\section{Quasi exact solvability (QES)}
On the one hand, we have a Hamiltonian $H(\vec{x})=-\Delta_{d} + V(\vec{x})$ with
$\vec{x}$ $\epsilon$ $\mathbb{R}$$^{d}$ and the corresponding eigenvalue equation
$H\psi(\vec{x})=E\psi(\vec{x})$. On the other hand, we have a Lie Algebra
${\mathcal G}$ of dimension $n_{\mathcal G}$ realised in terms of operators
$J_{k}=J_{k}(\vec{y}, \vec{\nabla}_{\vec{y}})$, $\vec{y}$
$\epsilon$ $\mathbb{R}$$^{d}$,
$k=1,...,n_{\mathcal G}$ which preserves a $m$-dimensional
vector space ${\mathcal V}$ of smooth functions.

Now, if there exists a) a change of variables
$\vec{x}=\vec{x}(\vec{y})$, b) an element ${\mathcal{H}}(\vec{y})$ of the 
enveloping algebra of ${\mathcal G}$ as well as c) an invertible scalar function
$f(\vec{x})$ such that $H(\vec{x})=f(\vec{x}){\mathcal{H}}(\vec{y})
f^{-1}(\vec{x})|_{\vec{y}=\vec{y}(\vec{x})}$, then the Hamiltonian
$H$ preserves the vector space $f{\mathcal V}$:
$H f{\mathcal V} \subseteq f{\mathcal V}$.
As a consequence, the restriction of the operator $H$ to the vector 
space ${\mathcal V}$ leads to an algebraic equation
\cite{[tur]}. This equation
allows to construct $m$ eigenvectors algebraically. 
In this construction, the factor $f$ which is called ``gauge factor''
plays a crucial role. The above procedure is usually called an algebraisation
of the Schr\"odinger problem. 

In this contribution, we will put the main emphasis on quantum Hamiltonians $H$
which possess multiple, i.e. several algebraisations. More specifically, more than
one ``gauge factor'' $f$, say $f_a$, $a=1,2,3...$ exists. This allows to write 
$H(\vec{x})=f_a(\vec{x}){\mathcal{H}}(\vec{y}) 
f_a^{-1}(\vec{x})|_{\vec{y}=\vec{y}(\vec{x})}$ where $J_{k,a}$ are now the
generators of the Lie Algebra ${\mathcal G}_{a}$. Correspondingly,
the invariant vector spaces ${\mathcal V}_a$ of the generators $J_{k,a}$ can also
be of different nature.

\section{Hamiltonians with multiple algebraisations}

\subsection{The Lam\'e equation}
A natural example of a QES equation for which multiple algebraisations
exists is the Lam\'e equation:
\begin{equation}
\label{lame}
\left(-\frac{d^2}{dx^2}+N(N+1) k^2 sn^2(x,k)\right)\psi(x)=E\psi(x)
\end{equation}
where $sn(x,k)$ is the Jacobi elliptic function of period $4K(k)$ with $K(k)$ being the
complete elliptic integral of first type. The potential
in this problem is periodic with period $2K(k)$ and the parameter
$N$ determines the height of the potential.
For the equation (\ref{lame}), four different algebraisations ($a=1,2,3,4$)
exist. For the case of even $N=2n$, $n$$\epsilon$$\mathbb{N}$, the gauge factors are given 
by:
\begin{equation}
f_1=1 \ , \ f_2=sn(x,k)cn(x,k) \ , \ f_3=sn(x,k)dn(x,k) \ , \
f_4=dn(x,k)cn(x,k) \ .
\end{equation}
The relevant change of variable is $y(x)=sn^2(x,k)$. Then ${\mathcal H}_{a}=
f_a H f^{-1}_a$ can be written in terms of the operators
\begin{equation}
J_{+}=y^2\frac{d}{dy}-n_a y \ , \ J_0=y\frac{d}{dy}-\frac{n_a}{2} \ , \ 
J_{-}=\frac{d}{dy} \ ,
\end{equation}
where $n_1=n$, $n_2=n_3=n_4=n-1$ and
${\mathcal V}_1={\mathcal P}(n)$, $
{\mathcal V}_2={\mathcal V}_3={\mathcal V}_4={\mathcal P}(n-1)$.
Here $ {\mathcal P}(n)$ denotes the set of polynomials of degree
less or equal to $n$ in $y$. The four underlying algebras ${\mathcal G}_a$
are in this case equal to $sl(2)$.

\subsection{Many-body Hamiltonians}
In this section, we will discuss how the idea of multiple algebraisation
can be applied to many-body Hamiltoninas with $N$ degrees of freedom.
The Hamiltonians considered here are generalisations
of Olshanetsky-Perelomov- \cite{[1]} and Inozemtsev \cite{[2]}-type.
More specifically, the Hamiltonian is given by:
\begin{eqnarray}
H_N(\vec{x})=-\Delta_N&+&a(a-1)\sum^N_{j,k=1\atop{j\not=k}}
\left({\it P}(x_j+x_k)+
{\it P}(x_j-x_k) \right) \nonumber \\
 &+& 4b(b-1)\sum_{k=1}^{N} {\it P}(2x_k)+
c\sum_{k=1}^{N} {\it P}(x_k+i\beta) \ .
\end{eqnarray}
${\it P}(z,g_2,g_3)$ denotes the Weierstrass function
of parameters $g_2$, $g_3$ which are related to the periods on the real and imaginary axes of the complex
plane, $2\alpha$ and $2i\beta$, respectively. $a$, $b$, $c$ are
coupling constants. For later use, we recall the following identities:
(i) $\left(\frac{d {\it P}(z)}{d z}\right)^2=4  {\it P}^3(z)
-g_2  {\it P}(z)-g_3$ and
(ii) ${\it P}(x+i\beta)=e_3+(e_2-e_3)sn^2(\sqrt{e_1-e_3}x,k)$
with $k=\frac{e_2-e_3}{e_1-e_3}$.
The singularity of the Weierstrass function at the origin (
$ {\it P}(\varepsilon << 1)=
\frac{1}{\varepsilon^2}+O(\varepsilon^2)$)
renders the potential singular for $|x_k-x_l|\rightarrow 0$ such that the
eigenvalue problem can be restricted to the domain
$D=\{(x_1,x_2,...x_N)| \ 0 < x_1 < x_2 < ... < x_N < \alpha\}$.
The condition that the wave function vanishes on the boundary of $D$
leads to a discrete spectrum of $H_N$.

Before addressing the main result of \cite{[3]}, we give the following
definition of symmetric polynomials
\cite{[4]}:
\begin{equation}
\tau_1=z_1+z_2+...+z_N \ , \ \tau_2=z_1z_2+z_1z_3+...+z_{N-1}z_N \ , \ .... \ , \
\tau_N=z_1z_2z_3...z_N
\end{equation}
and the definition of the vector space
\begin{equation}
{\mathcal V}_{m}=span \{\tau_1^{l_1}\tau_2^{l_2}....\tau_N^{l_N} \ , \
\sum_{j=1}^{N}l_j \le m \} \ , \ dim ({\mathcal V}_{m})=C_m^{N+m}
\end{equation}
where $C$ denotes the combinatoric symbol.
It is well known \cite{[4]} that the space ${\mathcal V}_{m}$
is preserved by a set of operators realising the Lie Algebra $sl(N+1)$.\\
\\
{\bf \underline{Proposition 1}} : Let
\begin{eqnarray*}
&{\rm (i)}& \ \  \mu(\vec{x})=\prod_{j<k}^{N} \left[{\it P}(x_j+i\beta)-
{\it P}(x_k+i\beta)\right]^a \prod_{k=1}^{N} \left[{\it P}^{'}
(x_k+i\beta)\right]^b \nonumber \\
&{\rm (ii)}& \ \  z_k={\it P}(x_k+i\beta) \nonumber \\
&{\rm (iii)}& \ \   {\mathcal H}_N(\vec{z})=\mu^{-1}(\vec{x}) H_N (\vec{x}) 
\mu(\vec{x})|_{\vec{x}=
\vec{x}(\vec{z})} \ .
\end{eqnarray*}
(i) is the ``gauge factor'' and (ii) the new variable.
Then ${\mathcal H}_N(\vec{z})$ depends only on the symmetric variables $\tau_1$,..., $\tau_N$
and
$ {\mathcal H}_N(\vec{\tau}){\mathcal V}_m \subseteq {\mathcal V}_m$
 iff
\begin{equation}
c\equiv c_m= [2m +2a(N-1) +4b]\cdot [2m+1 +2a(N-1) +2b] \ , m\epsilon \
\mathbb{N} \ .
\end{equation}
If in the above expression for the coupling constant $c$ the parameter $m$
is an integer the total number of algebraically obtainable eigenvectors 
is $C_m^{N+m}$. 

The five parameter Hamiltonian $H(a,b,c,e_2,e_3)$ has well-known limits
for specific choices of the parameters as is denoted in Table 1.
%
%
\begin{table}[t]
\caption{Specific limits of the five parameter Hamiltonian
$H(a,b,c,e_2,e_3)$ }
\vspace{2mm}
\small
\begin{center}
\begin{tabular}{|c|c|c|c|c|c|c|}           
\hline
\raisebox{0mm}[4mm][2mm] prameters & $\#$ of algebraic eigenvectors
& Remark \\
\hline\hline
\multicolumn{1}{|l|}{\raisebox{0mm}[4mm]{$(a,b,c,e_2,e_3)$}} & $0$ &
\\[0.5mm]
\hline   
\multicolumn{1}{|l|}{\raisebox{0mm}[4mm]{$(a,b,c_m,e_2,e_3)$}} & $C_N^{N+m}$ & see
\cite{[3]}
\\[0.5mm]
\hline
\multicolumn{1}{|l|}{\raisebox{0mm}[4mm]{$(a,0,c_m,e_2,e_3)$}} & $C_N^{N+m}+
3C_N^{N-1+m}$ & see further text 
\\[0.5mm]
\hline
\multicolumn{1}{|l|}{\raisebox{0mm}[4mm]{$(0,0,c_m,e_2,e_3)$}} & $(4m+1)^N$ & 
$N$ decoupled Lam\'e equations 
\\[0.5mm]
\hline
\multicolumn{1}{|l|}{\raisebox{0mm}[4mm]{$(0,0,c_m,e_2,e_2)$}} & infinite & 
$N$ free oscillators
\\[0.5mm]
\hline
\end{tabular}
\vspace{-1mm}
\end{center}
\end{table}

It should be remarked that out of the $(4m+1)^N$ eigenvectors
for the case $a=b=0$ a total number of $C_N^{N+m}+3C_N^{N-1+m}$ are completely
symmetric under permutations of the coordinates and as such can be written
in terms of the $\tau$-variables. This number contrasts with the
number of algebraic states existing in \cite{[4]}. This gives a hint
to the fact that further algebraisations might be possible to construct 
by introducing
different ``gauge factors''. 

In \cite{[5]}, we introduced these additional factors of the form
\begin{equation}
\label{al}
\tilde{\mu}(\vec{z})=\prod_{k=1}^{N}(z_k-e_1)^{\nu_1}(z_k-e_2)^{\nu_2}(z_k-e_3)^{\nu_3}
\end{equation}
such that
$\tilde{{\cal H}}_N(\vec{z})=
\tilde{\mu}^{-1}(\vec{z}){\cal H}_N(\vec{z})\tilde{\mu}(\vec{z})$.
This then leads to the following\\
\\
{\bf \underline{Proposition 2}}: If the exponents $\nu_i$, $i=1,2,3$ in (\ref{al}) are chosen according to
$\nu_i=0$ or $\nu_i=\frac{1}{2}-b$, then $\tilde{{\cal H}}_N(\vec{z}) {\mathcal M}_{\tilde{m}} 
\subseteq {\mathcal M}_{\tilde {m}}$ where $\tilde{m} := m+(b-\frac{1}{2})n_f$.\\
\\
In the above definition of $\tilde{m}$, $n_f$ denotes the number of non-zero
values chosen for the exponents $\nu_i$, that is to say $n_f=0$ or $1$ or $2$ or $3$.
This proposition results into the existence of seven possible new algebraisations.
The algebraisations studied in \cite{[4]} correspond to the choice 
$n_f=0$. The choices $n_f=1$ or $n_f=2$ lead to three new algebraisations
according to the possible permutation of the non-trivial factor. Finally, the choice $n_f=3$
obviously leads to only one new algebraisation. 
In order to avoid singularities of all possible ``gauge factors'', we will restrict 
$b$ to $0 \le b < \frac{1}{2}$.

It should be remarked that the previous condition $m$$\epsilon$ $\mathbb{N}$ 
is now replace by the condition
$\tilde{m}$$\epsilon$ $\mathbb{N}$. Then, for fixed $b$ and $n_f$ the parameter $m$ defining the
coupling constant $c$ has to be chosen appropriately. Though $m$ is still quantised, the number
of possible values of the coupling constant $c\equiv c_m$ is considerably increased.
To finish this section, we investigate for which fixed values
of the coupling constants $(a,b,c,e_2,e_3)$ multiple algebraisations exist.
For generic values of $b$ one (resp. three) algebraisations occur
for the case when $\tilde m$ is an integer and $n_f = 0$
or $n_f=3$ (resp. $n_f=2$ or $3$).
For $b=0$
four algebraisations coexist corresponding either to
(a) $n_f=0$ (with invariant   subspace ${\mathcal M}_m$)
and $n_f=2$ (with invariant subspace   ${\mathcal M}_{m-1}$)
or to (b) $n_f=1$ and $n_f=3$.
Finally the case $b=1/6$  leads to two algebraisation with $n_f=0$
and $n_f=3$ with respective invariant spaces ${\mathcal M}_m$ and
${\mathcal M}_{m-1}$.

%
%

\subsection{Coupled channel equation}
Apart from studying many body equations with algebraic properties
also systems of coupled Schr\"odinger equations of the
form
\begin{equation}
H_{2\times 2}=-\frac{d^2}{dx^2} 1\!{\rm I}_2+\left(\begin{array}{cc}
V_{11}(x) & V_{12}(x)\\
V^*_{12}(x) & V_{22}(x)\\
\end{array}
\right)
\end{equation}
can be studied. In \cite{[6]} a classification of equations
of this type which admit multiple algebraisations was achieved.
Here, we will study one example which has a physical motivation.
The potential reads:
\begin{equation}
V=sn^2(x,k)\left(\begin{array}{cc}
A-2b & 0\\
0 & A+2b\\
\end{array}
\right)
+b\left(\begin{array}{cc}
1 & 0\\
0 & -1\\
\end{array}
\right)
+\theta sn(x,k) cn(x,k) \left(\begin{array}{cc}
0 & 1\\
1 & 0\\
\end{array}
\right)
\end{equation}
where $k$ $\epsilon$ $[0:1]$, $b$ $\epsilon$ $\mathbb{R}$, $m\epsilon$
 $\mathbb{N}$ and $A=\frac{k^{2}}{2}(4m^2+2m+1)$, $\theta=4b^2-k^4(4m+1)^2$.
Using again the change of variable $y=sn^2(x,k)$ and an appropriate
``gauge factor'' which is now a $2\times 2$ matrix, we were able
to show that the Hamiltonian $H_{2\times 2}$ has a quadruple algebraisation.
Here we present two of the algebraisations and refer the reader for the remaining
ones to \cite{[6]}:\\
\\
$\left(F^{-1} H F\right)\left(\begin{array}{cc}
{\mathcal P}(m-1)\\
{\mathcal P}(m)\\
\end{array}
\right) \subseteq 
\left(\begin{array}{cc}
{\mathcal P}(m-1)\\
{\mathcal P}(m)\\
\end{array}
\right)$, $F=\left(\begin{array}{cc}
sn & 0\\
0 & cn\\
\end{array}
\right)\left(\begin{array}{cc}
1 & \kappa_1\\
 0 & 1\\
\end{array}
\right)$
\\
$\left(G^{-1} H G\right)\left(\begin{array}{cc}
{\mathcal P}(m-1)\\
{\mathcal P}(m-1)\\
\end{array}
\right) \subseteq 
\left(\begin{array}{cc}
{\mathcal P}(m-1)\\
{\mathcal P}(m-1)\\
\end{array}
\right)$, $G=\left(\begin{array}{cc}
sn \ cn \ dn & 0\\
0 & dn\\
\end{array}
\right)\left(\begin{array}{cc}
1 & -\frac{y}{\kappa_1}\\
 0 & 1\\
\end{array}
\right)$\\
\\
where $\kappa_1=-\theta/(2b+k^2(1+4m))$.
Note that in suitable realisations (which we will not present here)
the operator $F^{-1} H F$ can be interpreted as an element of the 
Super-Lie Algebra $osp(2,2)$, while $G^{-1} H G$ is an element of the
Lie Algebra $sl(2)\times sl(2)$. This provides an example
of multiple algebraisations where the different algebraisations
are related to distinct Lie Algebras. 

The limit $k=0$ of these coupled equations is non-trivial
and leads to the following operator:
\begin{equation}
\label{htrig}
H_{2\times 2}=-\frac{d^2}{dx^2} 1\!{\rm I}_2+ 2b\left(\begin{array}{cc}
\cos^2(x)-\frac{1}{2} & \cos(x) \sin(x)\\
\cos(x) \sin(x) & \sin^2(x)-\frac{1}{2}\\
\end{array}
\right)   \ \ \ , \ \ \ x \in [0, 2N \pi] \ .
\end{equation}
In this limit, the dependence of the potential on the integer
$m$ disappears and the system becomes exactly solvable.
In order to construct the flag of invariant
 vector spaces, it is useful to introduce the vectors
\begin{equation}
E(p) 
\equiv
\left(\begin{array}{cc}
\cos {px\over N}\\
\sin {px\over N}
\end{array}\right)
\quad , \quad G(p) 
\equiv
\left(\begin{array}{cc}
- \sin {px\over N}\\
\cos {px\over N}
\end{array}\right)
\quad , \quad p = 0, \pm 1 , \pm 2 , \cdots 
\end{equation} 
It is easy to show that the following vector spaces 
are left invariant by the operator (\ref{htrig}):
\begin{eqnarray}
\label{35}
V_0 &=&{\rm Span} \lbrace E(N)\rbrace\ ,\ 
\tilde V_0 = {\rm Span} \lbrace G(N) \rbrace\nonumber\\
V_k &=& {\rm Span} \lbrace E(N+k), E(N-k)\rbrace \quad  \ \ \ , \ \ k=1,2,3...\nonumber\\
\tilde V_k&=& {\rm Span} \lbrace G(N+k), G(N-k)\rbrace 
\ \ , \ \ \quad k=1,2,3...
\end{eqnarray}
The operator  can be diagonalized on any of these vector spaces 
and leads to the following sets of eigenvalues. 
Using natural notations, we find
\begin{eqnarray}
\label{36}
\omega^2(N,k,\pm) &=& 
 1-2b + \frac{k^2}{N^2}\pm \sqrt{b^2 + 4 \frac{k^2}{N^2}}
   \ , \ k=0,1,2, \dots  \\
\tilde \omega^2(N,k,\pm) &=& \omega^2(N,k,\pm)\ , \ k=1,2,3 \dots
\end{eqnarray}

This is reminiscent to the situation in the elliptic Perelomov model:
the elliptic model is not solvable for generic values
of the parameters determining the period
but in the limits corresponding to
rational or trigonometric potentials the model becomes
completely solvable.

The above system is related to the
normal mode analysis of the soliton appearing in the two dimensional
Goldstone-model with periodic boundary condition of the space variable
\cite{[7]}.
We would like to point out that coupled-channel Hamiltonians
which posses a multiple algebraisation are rare, very specific
 and difficult to
construct but many of them are related to the normal mode analysis
of classical solutions (solitons or sphaleron) of simple field
theories. These algebraic properties can be used as testing ground
for more ambitious models.

\bigskip
{\bf Acknowledgement}
Y. B. gratefully acknowledges the Belgian FNRS for financial
support. B. H. was supported by an EPSRC grant. We thank the
organisers of the conference in Prague for their hospitality.

\end{document}